\documentclass[prb,aps,twocolumn,showpacs]{revtex4-1}
\usepackage{amsfonts,amsmath,bm,multirow}
\usepackage[OT4]{fontenc}

\usepackage{graphicx}
\usepackage{xcolor}
\usepackage{epstopdf}
\usepackage{dsfont}
\usepackage{physics}
\usepackage{hyperref}

\usepackage{pdfpages}
\makeatletter
\AtBeginDocument{\let\LS@rot\@undefined}
\makeatother

\hypersetup{
	colorlinks,
	linkcolor={blue!90!black},
	citecolor={blue!90!black},
	urlcolor= {blue!90!black}
}

\newcommand{\rr}{{\bm{r}}}
\newcommand{\kp}{{\vb*{k}\dotproduct\vb*{p}}}

\renewcommand{\cp}{{\mathrm{c.p.}}}

\newcommand{\kk}{\bm{k}}

\renewcommand{\qq}{\bm{q}}

\newcommand{\eps}{\epsilon}

\newcommand{\eyy}{\epsilon_{yy}}
\newcommand{\ezz}{\epsilon_{zz}}

\newcommand{\hc}{\mathrm{H. c.}}

\begin{document}

%
%

\title{Hole spin-flip transitions in a self-assembled quantum dot}
\author{Mateusz Krzykowski}
\email{Mateusz.Krzykowski@pwr.edu.pl}
\author{Krzysztof Gawarecki}
\author{Pawe{\l} Machnikowski}
\affiliation{Department of Theoretical Physics, Wroc{\l}aw University of Science and Technology, Wybrze\.ze Wyspia\'nskiego 27, 50-370 Wroc{\l}aw, Poland}

\begin{abstract}
In this work, we investigate hole spin-flip transitions in a single self-assembled InGaAs/GaAs quantum dot. We find the hole wave functions using the 8-band $\kp$ model and calculate phonon-assisted spin relaxation rates for the ground-state Zeeman doublet. We systematically study the importance of various admixture- and direct spin-phonon mechanisms giving rise to the transition rates. We show that the biaxial and shear strain constitute dominant spin-admixture coupling mechanisms. Then, we demonstrate that hole spin lifetime can be increased if a quantum dot is covered by a strain-reducing layer.	Finally, we show that the spin relaxation can be described by an effective model.
\end{abstract}


\maketitle

\section{Introduction}
\label{sec:intr}

Dynamics of a carrier spin in semiconductor quantum dots (QDs), as well as other semiconductor nanostructures, is a subject of active studies, both theoretical and experimental, due to potential implementations in fields of spintronics and quantum information processing \cite{loss98, recher00}. High-fidelity initialization \cite{kugler11, gawelczyk13}, control \cite{bonadeo98, godden12}, readout and storage \cite{heiss07} of the information encoded in the spin are essential for future applications. These can be achieved with hole spin due to its relatively long coherence time \cite{huthmacher18}, which is related to significantly weaker hyperfine interaction as compared to the electron case \cite{fischer08, brunner09, vidal16}. However, spin life- and coherence times can be limited by the coupling to phonon bath, leading to the loss of information to the environment \cite{khaetskii00, khaetskii01, climente13}.

The channels of phonon-induced spin-flip can be divided into two classes \cite{khaetskii01, Mielnik-Pyszczorski2018b}. The first one contains various admixture mechanisms resulting from the spin-orbit coupling (SOC). Hence, a carrier state with some (dominant) spin orientation has also an admixture of the opposite spin. In consequence, the coupling to a phonon bath can lead to spin-flip transitions between such states \cite{khaetskii00, khaetskii01}. The second class of mechanisms results from the direct spin-phonon coupling. In this case, the displacement field related to phonons lowers the symmetry, leading to the spin relaxation in the presence of the spin-orbit coupling \cite{pikus84, Roth1960,khaetskii01,Mielnik-Pyszczorski2018b}.

The processes of hole spin-flip transition in a QD due to the mechanisms described above were widely studied \cite{woods04,lu05,bulaev05a,heiss07,climente13,wei12,Wei2014,Wei2014a,Segarra2015,Gawarecki2019a}. It has been shown that phonon coupling via piezoelectric field (PZ) is more important at small and moderate Zeeman splittings, while deformation potential (DP) coupling becomes dominate for larger splittings. A detailed comparison of various spin relaxation channels was presented in Ref.~\onlinecite{climente13}. Therein the problem was studied within a $4$-band $\kp$ model, neglecting structural strain, and combined with the approximation of parabolic potentials.
It is known, however, that strain provides channels of spin mixing \cite{Trebin1979} which can significantly affect the spin relaxation \cite{lu05}. In fact, recent results show an important contribution to the spin relaxation rates coming from the structural shear strain in a self-assembled QD \cite{Gawarecki2019a}.

In this paper, we systematically study the importance of various phonon-induced spin-flip transition mechanisms for a hole confined in InGaAs/GaAs quantum dot system. The hole states are calculated using the full 8-band $\kp$ method for a realistic geometry of the system. The structural strain distribution is accounted for within the continuous elasticity approach. The hole is subject to external magnetic field applied parallel to the growth direction as well as coupled to acoustic phonon reservoir via deformation potential and piezoelectric field. We show that the effect of spin-admixture mechanisms (dominant at low and moderate magnetic fields) coming from the shear- and biaxial strain, can be limited by the presence of a strain-reducing layer. Finally, we show that transitions via the spin-admixture channels can be accounted for using an effective model with Gaussian-like heavy-hole wave functions.

The paper is organized as follows. In Sec.~\ref{sec:model}, we describe the QD geometry and discuss the model used in the calculations. In Sec.~\ref{sec:results}, we present the results for various spin-flip transition mechanisms and introduce an effective model describing the relaxation due to spin-admixture effects. We conclude the article in Sec. \ref{sec:conclusion}.

\section{Model}
\label{sec:model}

\begin{figure}[tb]
	\begin{center}
		\includegraphics[width=70mm]{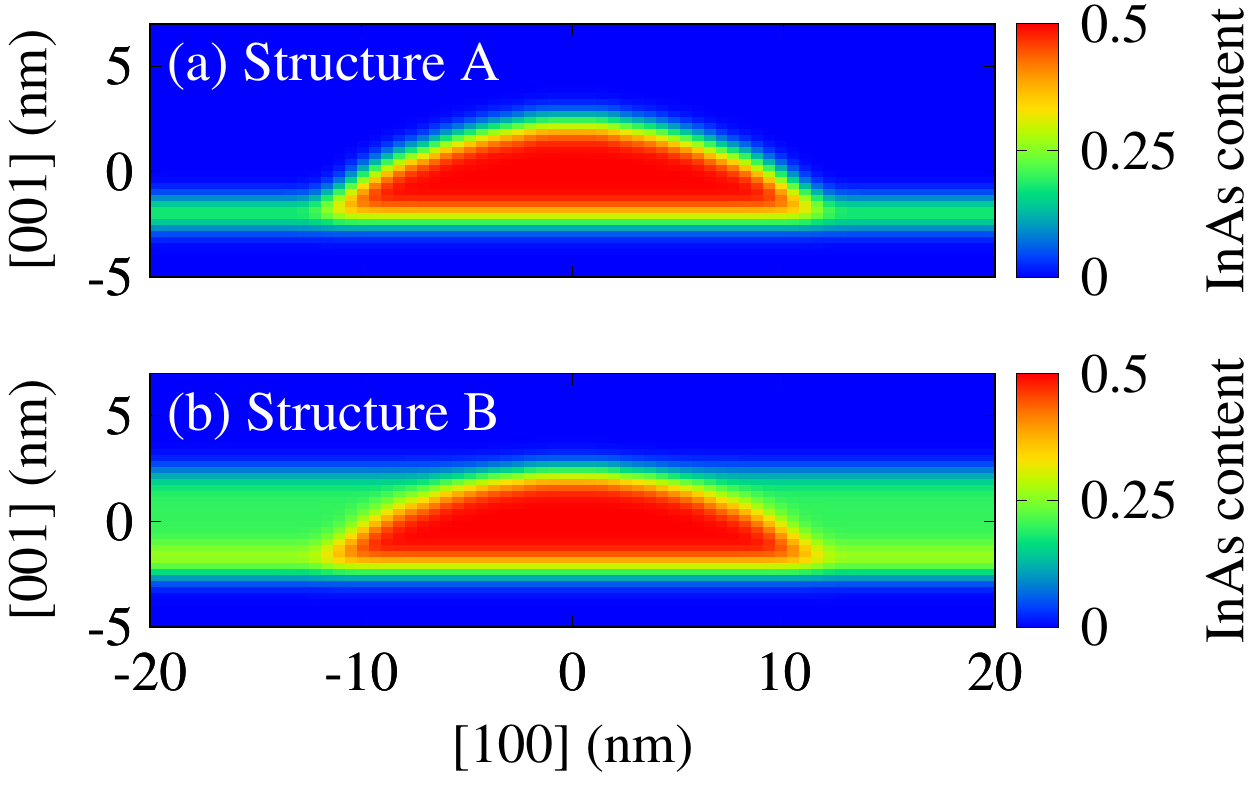}
	\end{center}
	\caption{\label{fig:comp}(Color online) In$_x$Ga$_{1-x}$As distribution in the system for the single QD (a) and the QD capped by SRL (b).
	}
\end{figure}

We consider a single, self-assembled QD of In$_{0.5}$Ga$_{0.5}$As/GaAs (the structure A, Fig.~\ref{fig:comp}a). The material intermixing is simulated by a Gaussian blur of the composition with the standard deviation of $0.6$~nm. We model the dot as a lens-shaped structure with a base radius of $21a$, height of $7a$, and $a$ thick wetting layer, where $a$ is the GaAs lattice constant. In the case of the structure B (see Fig.~\ref{fig:comp}b), the QD is capped by a In$_{0.2}$Ga$_{0.8}$As/GaAs strain-reducing layer (SRL) of constant thickness. Such layers are often utilized to tune QD emission to a desired range \cite{Nishi1999,Goldmann2013,Mrowinski2019}. In this paper, we use a SRL to soften strain at the interfaces.

The strain field caused by the lattice mismatch of InAs and GaAs materials is calculated using the continuous elasticity approach \cite{pryor98b}. The piezoelectric potential (inevitable in zinc-blende structure in the presence of shear strain) is calculated including the polarization up to the second order in strain tensor elements \cite{Bester06b}, where we use parameters from Ref.~\onlinecite{Caro2015}. Hole wave functions are obtained using the 8-band $\kp$ model \cite{bahder90,Winkler2003}. We incorporated magnetic field according to the gauge-invariant scheme described in Ref. \cite{andlauer08}. The computational domain is discretized on a cartesian mesh of $a \times a \times (a/2)$ cell size. The $\kp$ model and its implementation is described in detail in the Appendix of Ref.~\onlinecite{Gawarecki2018a}. Moreover, we supplement the Hamiltonian by the additional strain terms \cite{Trebin1979}
\begin{equation*}
\begin{split}
H_{\mathrm{str},\mathrm{6c8v}} & = i \sqrt{3} C_{2} \left[ T_{x} \epsilon_{yz} + \cp \right]\\
H_{\mathrm{str},\mathrm{6c7v}} & = - i \frac{1}{\sqrt{3}} C_{2} \left[ \sigma_{x} \epsilon_{yz} + \cp \right],
\end{split}
\end{equation*}
where 6c, 8v and 7v refer to the conduction- and valence band blocks, $\eps_{ij}$ are strain tensor components $\sigma_i$ are the Pauli matrices, $T_i$ are matrices connecting the $j=1/2$ and $j=3/2$ representations \cite{Winkler2003}, and $\cp$ denotes cyclic permutations. Due to lack of available experimental data for InAs, we assume the value of $C_{2}$ parameter for In$_x$Ga$_{1-x}$As as $C_2(x) = 0.4 \, E_{\mathrm{g}}(x)[E_{\mathrm{g}}(x) + \Delta(x)]/\Delta(x)$, where $E_{\mathrm{g}}$ is the energy gap, $\Delta$ is the spin-orbit parameter, and $0.4$ was extracted from the experimental data for GaAs \cite{dyakonov86a}.

We also take into account terms (in valence band blocks) which are proportional to the (bi-)axial strain and $\kk$ components \cite{Trebin1979,Winkler2003}
\begingroup
\allowdisplaybreaks
\begin{subequations}
	\begin{align*}
	H^{\mathrm{(k)}}_{\mathrm{str},\mathrm{8v8v}} &= [ C_4 (\eyy - \ezz) k_x ] J_x + \cp,\\
	H^{\mathrm{(k)}}_{\mathrm{str},\mathrm{8v7v}} &= \frac{3}{2} [ C_4 (\eyy - \ezz) k_x ] T_x^\dagger + \cp,\\
	H^{\mathrm{(k)}}_{\mathrm{str},\mathrm{7v7v}} &= [ C_4 (\eyy - \ezz) k_x ] \sigma_x + \cp,
	\end{align*}
\end{subequations}
\endgroup
where $J_i$ are matrices of angular momentum ($j=3/2$). The form of $H^{\mathrm{(k)}}_{\mathrm{str},\mathrm{7v7v}}$ was derived from the table of irreducible tensor components of $T_\mathrm{d}$ point group given in Ref. \cite{Winkler2003}.
There are significant discrepancies in the reported values of $C_4$ parameter. While the empirical pseudopotential method (EPM) gives $C_{4}[\mathrm{InAs}] = 2.9$~eV\AA, and $C_{4}[\mathrm{GaAs}] = 3.2$~eV\AA; the results of $sp^3$ tight-binding (TB) model suggest $C_{4}[\mathrm{InAs}] = 7.0$~eV\AA, and $C_{4}[\mathrm{GaAs}] = 6.8$~eV\AA \cite{Silver1992}. Although in the present paper we utilize the EPM parameterization, we note that the values from the TB model lead to considerably stronger spin mixing.

In Ref. \cite{Jancu2005}, it has been shown that the off-diagonal spin-orbit parameter $\Delta^{-}$ significantly contributes to the Dresselhaus coupling. Such a parameter is inherently present in the full $14$-band $\kp$ Hamiltonian \cite{Cardona1988}. In this work, we represent its influence (within the $8$-band $\kp$ Hamiltonian) perturbatevely, where we took $\Delta^-[\mathrm{InAs}] = -0.05$~eV, and $\Delta^-[\mathrm{GaAs}] = -0.17$~eV \cite{Jancu2005}.\\

The full Hamiltonian of the system can be written as \cite{grodecka05a}
\begin{equation*}
\begin{split}
H =& \sum_{n} E_n h^{\dagger}_n h_n + \sum_{\lambda, \qq} \hbar \omega_{\lambda, \qq} b_{\lambda, \qq}^\dagger b_{\lambda, \qq} + \sum_{ij} V_{ij} h^{\dagger}_i h_j ,
\end{split}
\end{equation*}
where $E_n$ describes the energy of the $n$-th state and $h^{(\dagger)}_n$ is the related annihilation (creation) operator. The second term accounts for the phonon bath, where $\lambda \in \qty{\mathrm{l, t_1, t_2}}$ denotes the acoustic phonon branch (single longitudinal and two transversal modes respectively), $\qq$~is a wave vector, $\hbar \omega_{\lambda, \qq}$ is a phonon mode energy and $b_{\lambda, \qq}^{(\dagger)}$ is the annihilation (creation) operator of the mode. We assume the linear dispersion $ \omega_{\lambda, \qq} = c_\lambda q$ with a branch-dependent speed of sound $c_l = 5150$~m/s, and $c_{t1/t2} = 2800$~m/s \cite{Levinshtein1999}. The last term accounts for the hole-phonon interaction
\begin{equation*}
V_{ij} = \int d^3 \bm{r} \bm{\Psi}^{\dagger}_i (\bm{r}) \qty [ H^{(\mathrm{ph})}_{\mathrm{DP}}(\rr) + V^{(\mathrm{ph})}_{\mathrm{PZ}}(\rr) ] \bm{\Psi}_j (\bm{r}),
\end{equation*}
where $\bm{\Psi}_{i}(\bm{r})$ is a wave function of the $i$-th hole state in the form of eight-component pseudo-spinors \cite{Winkler2003,eissfeller12}, while $H^{(\mathrm{ph})}_{\mathrm{DP}}$, and $V^{(\mathrm{ph})}_{\mathrm{PZ}}$ represent the carrier-phonon couplings via deformation potential and piezoelectric field respectively.

The deformation potential coupling is described by the Bir-Pikus Hamiltonian supplemented with the $C_{2}$-strain terms \cite{Trebin1979,bahder90, Winkler2003},
\begin{equation*}
\begin{split}
H^{(\mathrm{ph})}_{\mathrm{DP}}(\rr) &= a_{\mathrm{c}} \Tr{\hat{\epsilon}(\bm{r})} \mathbb{I}^{\mathrm{(6c)}} + a_{\mathrm{v}} \Tr{\hat{\epsilon}(\bm{r})} \mathbb{I}^{\mathrm{(8v+7v)}} \\
&\phantom{=} - b_{\mathrm{v}} \qty[ \qty( {J^{\mathrm{(8v)}}_{x}}^2 - \frac{1}{3} {J^{\mathrm{(8v)}}}^2 )\epsilon_{xx}(\bm{r}) + \cp ] \\
&\phantom{=} - \frac{d_{\mathrm{v}}}{\sqrt{3}} \left[ 2 \qty{J^{\mathrm{(8v)}}_{x},J^{\mathrm{(8v)}}_{y}} \epsilon_{xy}(\bm{r}) + \cp \right] \\
&\phantom{=} -3 b_{\mathrm{v}} \qty[ \qty(T^{\mathrm{(7v8v)}}_{xx} + \hc ) \epsilon_{xx}(\bm{r}) + \cp ] \\
&\phantom{=} -2 \sqrt{3} d_{\mathrm{v}} \left[ \qty(T^{\mathrm{(7v8v)}}_{xy} + \hc ) \epsilon_{xy}(\bm{r}) + \cp \right] \\
&\phantom{=} + \sqrt{3} C_{2} \left[ \qty(i T^{\mathrm{(6c8v)}}_{x} + \hc ) \epsilon_{yz}(\bm{r}) + \cp \right]\\
&\phantom{=} - \frac{1}{\sqrt{3}} C_{2} \left[ \qty(i \sigma^{\mathrm{(6c7v)}}_{x} + \hc ) \epsilon_{yz}(\bm{r}) + \cp \right],
\end{split}
\end{equation*}
where $a_{\mathrm{c}}$, $a_{\mathrm{v}}$, $b_{\mathrm{v}}$, $d_{\mathrm{v}}$ are deformation potentials, $\hat{\epsilon} (\bm{r})$ is the (phonon-induced) strain-tensor field; $\mathbb{I}$ is an identity matrix and $\sigma, J, T$ are matrices used for the invariant expansion of the Hamiltonian with the superscripts referring to the band blocks
\cite{Supplementary}.
We take GaAs values for all deformation potentials and $C_{2}$ parameter. To obtain $H^{(\mathrm{ph})}_{\mathrm{DP}}$ in the representation of phonon normal modes, we perform the expansion $$\epsilon_{ij}(\bm{r}) = \sum_{\lambda, \qq} \epsilon^{(\bm{q},\lambda)}_{ij} e^{i \qq \rr},$$ with the coefficients \cite{grodecka05a}
\begin{equation*} \label{modes}
\eps^{(\bm{q},\lambda)}_{ij} = - \frac{1}{2} \sqrt{\frac{\hbar}{2 V \!\rho \omega_{\lambda ,\qq}}} \left( \hat{e}_{\qq \lambda, i} q_j + \hat{e}_{\qq \lambda, j} q_i \rule{0pt}{9pt} \right) \left( b_{-\qq,\lambda}^\dagger + b_{\qq,\lambda} \right),
\end{equation*}
where $\rho = 5350$~kg/m$^3$ \cite{Levinshtein1999} and $V$ denote density and volume in the bulk crystal, respectively, and $\hat{e}_{\qq\lambda, i}$ is the $i$-th component of the polarization unit vector.

The coupling via piezoelectric field potential is given by $ V^\mathrm{(ph)}_\mathrm{PZ} = e \phi(\bm{r}) \mathbb{I}^{\mathrm (6c+8v+7v)},$
where $e$ is the elementary charge, $\phi(\bm{r})$ is the phonon-induced electrostatic potential given by \cite{uenoyama90b,grodecka05a}
\begin{equation}
\phi(\bm{r}) = i \frac{2 d_{\mathrm{14}}}{\varepsilon_0 \varepsilon_r} \sum_{\qq,\lambda} \frac{1}{q^2} \left( q_x \epsilon^{(\bm{q},\lambda)}_{yz} + \cp \right) e^{i \bm{q} \bm{r}},
\end{equation}
where $\varepsilon_r = 12.4$ is the relative dielectric constant (here assumed equal to the bulk GaAs value \cite{Blakemore1982}), $d_{14} = -0.16$~C/m$^2$ is the element of piezoelectric tensor (in a zinc-blende crystal only one component is linearly independent) for GaAs \cite{Adachi1992}.

The interaction Hamiltonian can be expressed by $$V_\mathrm{int} = \sum_{\lambda,\qq} \mathcal{V} (\qq,\lambda) e^{i \qq \rr},$$
where $\mathcal{V} (\qq,\lambda)$ contains $H^{(\mathrm{ph})}_{\mathrm{DP}}$ and $V^{(\mathrm{ph})}_{\mathrm{PZ}}$ for a single phonon mode ($\qq,\lambda$).
We calculate the phonon-induced relaxation rates using the Fermi golden rule. The rate between the $i$ and $j$ states is $\Gamma_{ij} = 2 \pi R_{ijji}\qty(\frac{E_{i} - E_{j}}{\hbar})$, where $R_{ijji}(\omega)$ is the phonon spectral-density given by \cite{grodecka05a}
\begin{equation*}
R_{ijji}(\omega) = \frac{1}{\hbar^2} \sum_{\lambda, \qq} \abs{\matrixel{\psi_i}{\mathcal{V} (\qq,\lambda) e^{i \qq \rr}}{\psi_j}}^2
\delta \left(\omega - \omega_{\lambda, \qq} \right),
\end{equation*}
where we assume absolute zero temperature.

\section{Results}
\label{sec:results}

\subsection{Full model} \label{sec:numerical}

In this section, we analyze hole spin-flip transitions due to two distinct classes of mechanisms. The first class is induced by
band-off-diagonal terms in the multiband carrier-phonon interaction Hamiltonian. 
As discussed in detail in Ref.~\cite{Mielnik-Pyszczorski2018b}, when the multi-band Hamiltonian is reduced to an effective two-band model (describing the two heavy-hole subbands in the present case) via L\"owdin perturbational decoupling, such terms lead to direct spin-phonon couplings in the effective Hamiltonian. 
The second mechanism relies on the band-off-diagonal terms unrelated to phonons that express various spin-orbit couplings. As a result, the predominantly heavy-hole state with a certain nominal spin orientation has contributions (admixtures) of states with the opposite spin. Therefore, states with nominally opposite spins may be coupled via
spin-conserving phonon terms (stemming from the diagonal elements in the $\kp$ representation). Although L\"owdin elimination is of less practical use for holes than for electrons, the relation between the location of the phonon term in the $\kp$ Hamiltonian and the form of the resulting
effective term still holds in principle and allows one to classify the numerous spin relaxation channels. Therefore, we will use the terms \textit{spin-phonon} and \textit{admixture} to label spin-flip mechanisms in the following discussion even though we study the full multi-band $\kp$ model.   

\begin{figure}[!tb]
\begin{center}
\includegraphics[width=80mm]{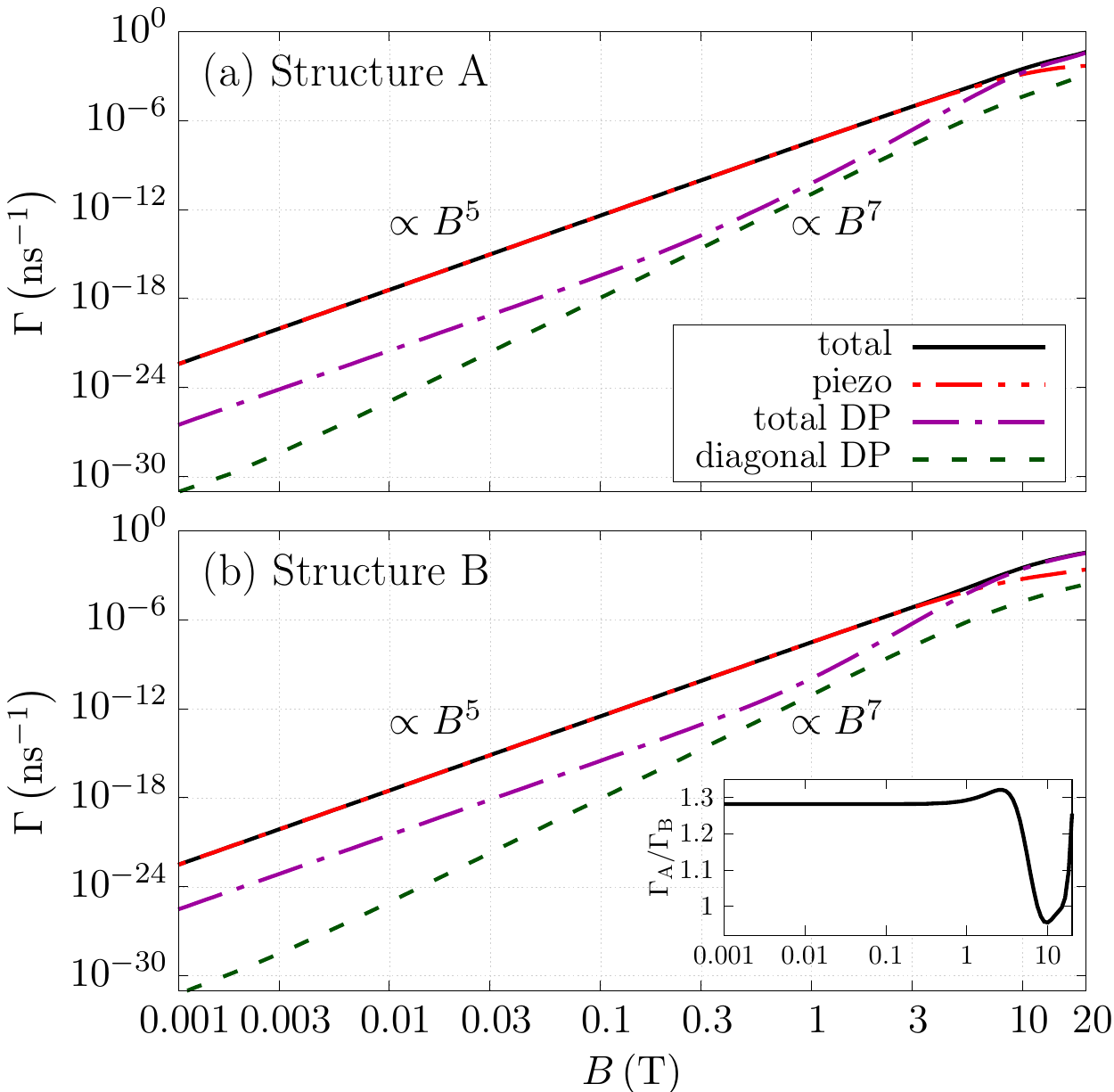}
\end{center}
\caption{\label{fig:phonon}{(Color online)}
Phonon-induced spin-flip relaxation rate in the lowest-energy Zeeman doublet as a function of axial magnetic field for the QD structures A and B. Solid black line indicates total rate while dashed colored lines indicate various contributions. The lines are ordered depending on their contribution to the considered process from greatest to lowest at low magnetic fields. The inset in panel (b) presents the ratio between total relaxation rates in the considered structures.
}
\end{figure}

In the Fig.~\ref{fig:phonon}, we present spin-flip relaxation rates due to the interaction with the acoustic phonon reservoir as a function of the magnetic field $B$ applied parallel to the growth axis [001]. Various lines in the figure correspond to particular contributions to the interaction Hamiltonian. The rates are given for the two QD structures (see~Fig.~\ref{fig:comp}) differing by the presence or absence of the SRL and therefore strain distribution in the system. For low magnetic fields, the coupling via piezoelectric field dominates, however around $10$~T for the structure A and $7$~T for the structure B this effects starts to saturate and the coupling via the deformation potential becomes dominant.
This effect is caused by different power dependencies of various spin-flip channels on magnetic field \cite{Climente2013a,Mielnik-Pyszczorski2018b}. The coupling via phonon-induced piezoelectric field exhibits $\propto B^5$ behavior. On the other hand, the coupling via deformation potential contains terms increasing like $\propto B^5$, $\propto B^7$, and even $\propto B^9$, which is clearly visible in the line slopes in the logarithmic scale. The $\propto B^9$ contribution is related to some of the $d_{\mathrm{v}}$ shear-strain off-diagonal terms in $H^{(\mathrm{ph})}_{\mathrm{DP}}$. Since in our approach $V^{(\mathrm{ph})}_{\mathrm{PZ}}$ is diagonal, it is spin-conserving and gives rise to the spin-flip relaxation due to the admixture mechanisms only. On the other hand, for the coupling via deformation potential, the off-diagonal terms clearly dominate over the diagonal part. In consequence, for high magnetic fields the direct spin-phonon coupling is the most significant class of mechanisms.

We have compared the relaxation rates for the structures with and without SRL [see the inset in Fig.~\ref{fig:phonon}(b)]. In the presence of SRL, the relaxation due to strain-related spin admixtures slows down. On the other hand, in the structure B, the direct spin-phonon mechanisms are stronger. In consequence, SRL increases the spin lifetime for weak and moderate magnetic fields, where the admixture mechanisms dominate.

\begin{figure}[!tb]
	\begin{center}
		\includegraphics[width=80mm]{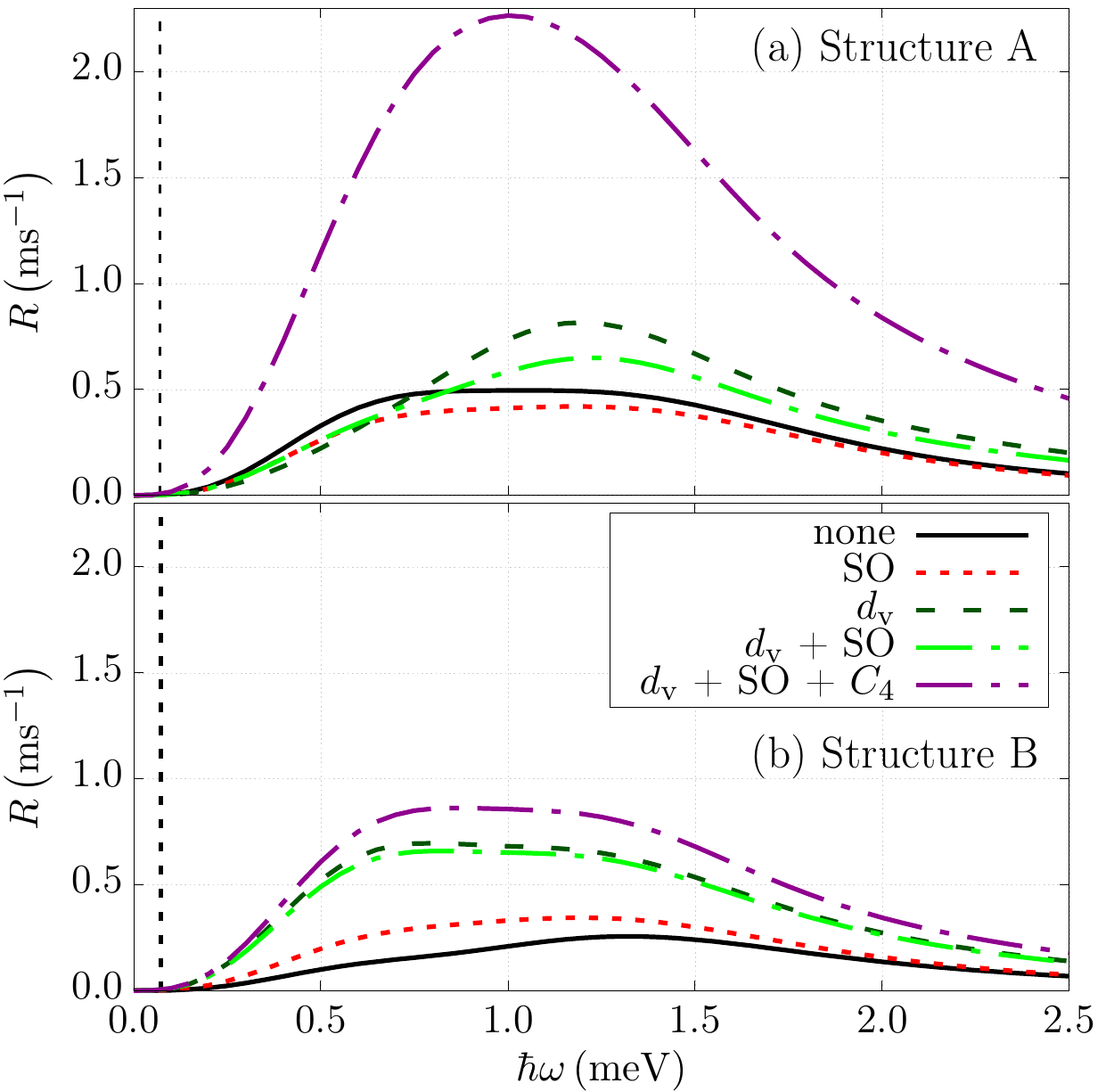}
	\end{center}
	\caption{\label{fig:admixture}{(Color online)}
		Phonon spectral density at $B = 1$~T, for the QD structures A and B, representing the absence and presence of a SRL, respectively. Solid black line represents the background related to various contributions, with all explicit Hamiltonian terms being turned off. Colored dashed lines represent cases of explicit terms being taken into account. The vertical line corresponds to the energy difference $E_{2} - E_{1}$ between the states in Zeeman doublet.
	}
\end{figure}

We have investigated the importance of various spin admixture mechanisms to the phonon-assisted spin-flip relaxation rate. This could be done by artificially turning on and off relevant explicit terms in the 8-band $\kp$ Hamiltonian. However, this may strongly affect the hole g-factor, hence the resulting relaxation rates correspond to different energies, making them hard to compare. Instead, we studied the relevant spectral densities (Fig.~\ref{fig:admixture}). To remove all of the contributions coming from the direct spin-phonon mechanisms, we took into consideration only the coupling via the piezoelectric field. We performed the simulations for both considered QD systems at $B = 1$~T (where, according to Fig.~\ref{fig:phonon}, the coupling via PZ field is the dominant contribution).

In the case of the QD without the SRL [Fig.~\ref{fig:admixture}(a)], the dominant contribution is the biaxial strain (terms proportional to $C_4$). Another significant contributions come from the shear strain in the valence band (terms proportional to $d_{\mathrm{v}}$ in Bir-Pikus Hamiltonian) as well as the background related to the remaining contributions, such as Rashba effect coming from structure inversion asymmetry (SIA), represented in the Fig.~\ref{fig:admixture} as solid black line. These results are consistent with Ref.~\onlinecite{Gawarecki2018b}, where $d_{\mathrm{v}}$ shear-strain terms were shown as one of the most important contributions determining the hole $s$-$p$ coupling. The contribution from the Dresselhaus bulk inversion asymmetry (BIA) spin-orbit coupling is relatively small for the structure A and it interferes destructively with the background.

In the case of the structure B [where SRL is included, Fig.~\ref{fig:admixture}(b)] the dominant $C_4$ contribution is significantly quenched, leading to an almost threefold reduction in spectral density. The Dresselhaus term remains mostly unchanged while the background contributions are reduced. This is to be expected for a low-strain regime, where the Dresselhaus effect can be the dominant spin-admixture-related relaxation channel \cite{climente13}. 
In contrast to the electron case \cite{Mielnik-Pyszczorski2018b}, for both considered structures the effect of off-diagonal terms linear in momentum and strain (in $H_{\mathrm{6c8v}}$ and $H_{\mathrm{6c7v}}$ blocks) is negligible. Furthermore, the influence of $C_{2}$-strain is small and is included in the background for clarity.

\subsection{Effective model} \label{sec:effective}

The hole phonon-assisted spin-flip relaxation rates can be accounted for using wave functions obtained from the effective model with empirical parameters fitted to the $\kp$ simulation data. We use heavy-hole Gaussian wave functions and a simple hole Hamiltonian based on the Fock-Darwin model, supplemented by additional terms accounting for the spin-orbit interaction \cite{Gawarecki2018b}.

\begin{table}[!t]
	\caption{Effective Hamiltonian and Gaussian wave functions parameters used in the effective model calculations.} \label{tab:params}
	\begin{ruledtabular}
		\begin{tabular}{lclc}
			$\Delta V_0^{(p)}$ & 15.92~meV &
			$V_{pp}^{(so)}$ & 3.132~meV \\
			$V_a$ & 5.474~meV &
			$V_{sp}^{(so)}$ & 164.1~$\mu$eV \\
			$g_s$ & 1.428 &
			$\alpha_s$ & 1.999~$\mu$eV/T$^2$ \\
			$g_p$ & 4.633 &
			$\alpha_p$ & 3.479~$\mu$eV/T$^2$\\
			$W$ & -0.288~meV/T & & \\ \hline
			$l_{z}$ & 15.69~nm & $l_{p}$ & 55.79~nm
		\end{tabular}
	\end{ruledtabular}
\end{table}

Let us consider the basis of $s$- and $p$-type states $\qty{ \ket{0 \Uparrow}, \ket{0 \Downarrow}, \ket{+1 \Uparrow}, \ket{+1 \Downarrow}, \ket{-1 \Uparrow}, \ket{-1 \Downarrow} }$, where the indices correspond to the axial projections of envelope and band angular momentum ($j_{z} = \pm 3/2$) respectively. The wave functions used for the calculations are Gaussians expressed in cylindrical coordinate system
\begin{equation*}
\begin{split}
\bm{\psi}^{(\nu)}_0(r,z) &= \frac{1}{\sqrt{\pi^{\frac{3}{2}} l_p^2 l_z}} \exp(-\frac{r^2}{2 l_p^2} - \frac{z^2}{2l_z^2}) \bm{\chi}_\nu, \\
\bm{\psi}^{(\nu)}_{\pm 1}(r,\phi,z) &= \frac{r}{\sqrt{\pi^{\frac{3}{2}} l_p^4 l_z}} \exp(- \frac{r^2}{2l_p^2} - \frac{z^2}{2l_z^2}) \exp(\pm i \phi) \bm{\chi}_\nu,
\end{split}
\end{equation*}
where $ \bm{\chi}_\nu$ is a spinor corresponding to the axial projection of the band angular momentum and $l_p$ and $l_z$ describe the spatial extension of the wave function in the $xy$ plane and along the $z$ axis, respectively. The effective Hamiltonian written in the considered basis is given by \cite{Gawarecki2018b}
\begin{equation*} \label{eff_hamiltonian}
	\begin{split}
		H_\mathrm{eff} =& \phantom{+} \Delta V_0^{\mathrm{(p)}} (\ketbra{1}{1} + \ketbra{-1}{-1}) \otimes \mathbb{I}_2 \\
		&+ \frac{1}{2} V_{\mathrm{a}} L_z \otimes \sigma_z + W B_z L_z \otimes \mathbb{I}_2 \\
		&+ \frac{1}{2} \mu_\mathrm{B} \qty[ g_{\mathrm{s}} \ketbra{0}{0} + g_{\mathrm{p}} \qty( \ketbra{1}{1} + \ketbra{-1}{-1} ) ] B_z \otimes \sigma_z \\
		&+ V_{\mathrm{pp}}^{\mathrm{(so)}} (\ketbra{1}{-1} + \ketbra{-1}{1}) \otimes \mathbb{I}_2 \\
		&+ V_{\mathrm{sp}}^{\mathrm{(so)}} [( \ketbra{0}{-1} \otimes \ketbra{\Uparrow}{\Downarrow} - \ketbra{0}{1} \otimes \ketbra{\Downarrow}{\Uparrow}) + \hc] \\
		&+ \qty[\alpha_{\mathrm{s}} \ketbra{0}{0} + \alpha_{\mathrm{p}} (\ketbra{1}{1} + \ketbra{-1}{-1})] B_z^2 \otimes \mathbb{I}_2,
	\end{split}
\end{equation*}
where $\Delta V_0^{\mathrm{(p)}}$ is a bare energy difference between $s$- and $p$-type states at $B = 0$, $V_{\mathrm{a}}$ is a parameter related to the anisotropy, $W$ is a parameter accounting for the influence of the envelope angular momentum, $\mu_\mathrm{B}$ is the Bohr magneton, $g_{\mathrm{s/p}}$ are effective g-factors for $s$- and $p$-type states respectively, $V_{\mathrm{pp}}^{\mathrm{(so)}}$ corresponds to the spin-orbit coupling for the $p$-type states, $V_{\mathrm{sp}}^{\mathrm{(so)}}$ describes coupling between $s$- and $p$-type states involving change both envelope and band angular momenta, $\alpha_{\mathrm{s/p}}$ are diamagnetic parameters for $s/p$-type orbitals, $L_z$ is the operator of the axial component of the envelope angular momentum, $\sigma_z$ is the axial Pauli matrix and finally $\mathbb{I}_n$ is a identity matrix of order $n$. All of the parameters describing the effective Hamiltonian are fitted to the magnetic-field dependence of the energy levels obtained from $8$-band $\kp$ (see details in Refs.~\onlinecite{Gawarecki2018a, Gawarecki2018b}). Since, the fitting procedure gives only the absolute values of parameters, the relative phases of terms in $H_\mathrm{eff}$ are assumed.
\begin{figure}[!t]
	\begin{center}
		\includegraphics[width=80mm]{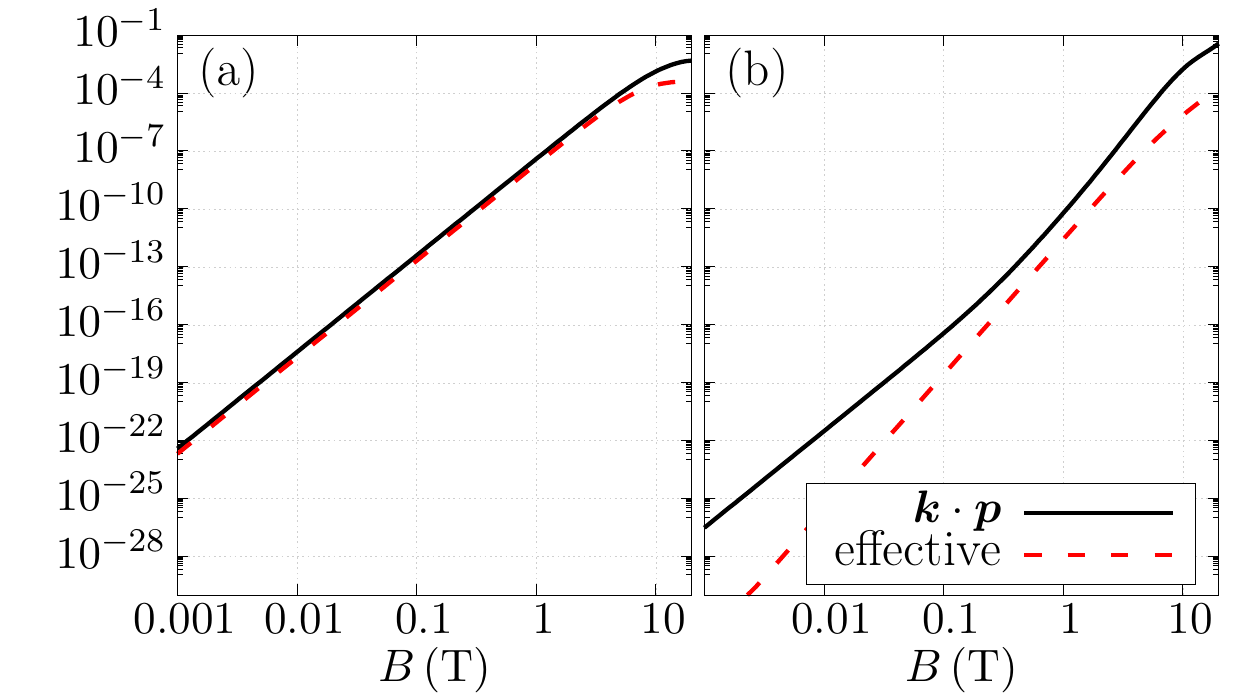}
	\end{center}
	\caption{\label{fig:gauss}(Color online) Phonon-induced spin-flip relaxation rate via piezoelectric effect (a) and deformation potential (b) couplings as a function of magnetic field $B$ for realistic $\kp$ calculations (solid black) and the effective model (dashed red).
	}
\end{figure}
We also neglected terms $\ketbra{0}{-1} \otimes \ketbra{\Downarrow}{\Uparrow} +\hc$ and $\ketbra{0}{1} \otimes \ketbra{\Uparrow}{\Downarrow} + \hc$ since they are not represented by any avoided crossing in target magnetic-field energy dependence \cite{Gawarecki2018b}. The parameters describing wave function spacial extension ($l_{p}$, $l_{z}$) are extracted from probability density maps at $B=0$. Finally, the effective Hamiltonian $6\times6$ matrix is diagonalized and relaxation rates are calculated using Fermi golden rule (with the same interaction Hamiltonian like in the full model).

We compare the values of relaxation rates obtained from the effective model and 8-band $\kp$ calculations for the QD without SRL (structure A). Fig.~\ref{fig:gauss}(a) presents the spin-flip relaxation rate via piezoelectric field alone. The results show a reasonable agreement, and the characteristic $\propto B^5$ dependence.
In the Fig.~\ref{fig:gauss}(b), we present similar comparison, but for the coupling via deformation potential alone. In this case, the results strongly disagree. In particular, the effective model does not reproduce $\propto B^5$ and $\propto B^9$ dependence regimes characteristic for direct spin-phonon coupling.

\section{Conclusions}
\label{sec:conclusion}

We have presented a theoretical study of the hole phonon-assisted spin-flip relaxation in a self-assembled QD systems. With wave functions found using $8$-band $\kp$ method, we have calculated relaxation rates related to the phonon coupling via the deformation potential and the piezoelectric field. In this framework, we have investigated the contributions coming from various channels belonging to the spin-admixture and direct spin-phonon classes of mechanisms. We have shown that the dominating spin-admixture terms come from the biaxial- and shear strain.
We have shown that (for low and moderate magnetic fields) the QD covered by a strain reducing layer offers significantly longer spin lifetime compared to the bare QD system. Finally, we have demonstrated that a relatively simple effective model gives a reasonable agreement to the $\kp$ simulation data.

\acknowledgments
This work was supported by the Diamentowy Grant Program of the Polish Ministry of Science and Higher Education No. 0062/DIA/2017/46 (M.K.) and from the Polish National Science Centre (NCN) under Grant
No. 2016/23/G/ST3/04324 (K.G., P.M.).
Part of the calculations have been carried out using resources provided by Wroc{\l}aw Centre for Networking and Supercomputing (\url{http://wcss.pl}), Grant No.~203.
We are also grateful to Micha{\l} Gawe{\l}czyk for sharing his implementation of the blur algorithm.

\bibliographystyle{prsty}
\bibliography{abbr,quantum,library.bib}

\setboolean{@twoside}{false}
\includepdf[pages={{},{},{},1}]{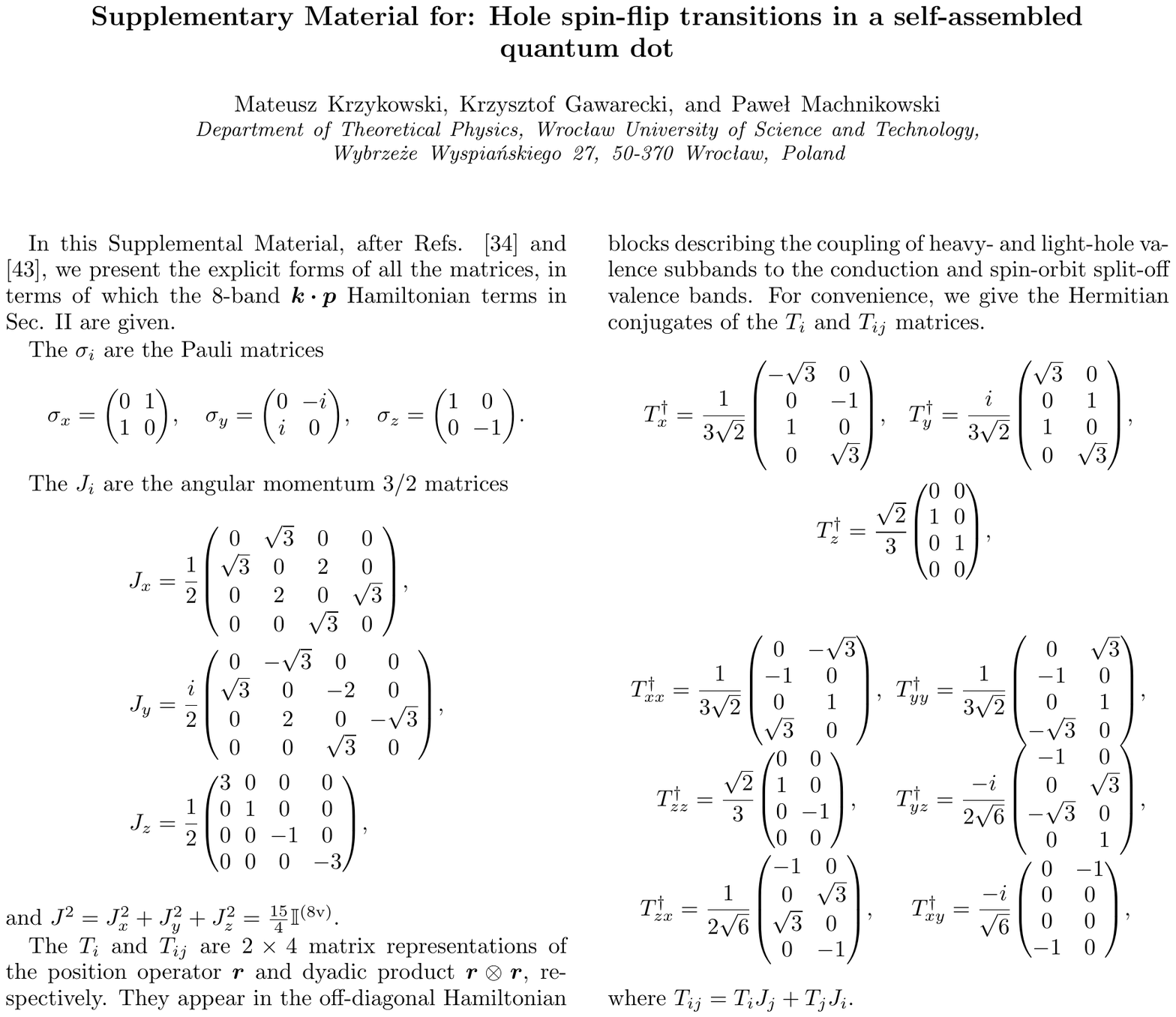}\AtBeginShipout\AtBeginShipoutDiscard

\end{document}